\documentclass[doublecol]{epl2}
\usepackage{amsmath}

\title{The Hawking temperature in the context of dark energy}

\author{Debashis Gangopadhyay\inst{1}\thanks{E-mail: \email{debashis@rkmvu.ac.in}} \and
 Goutam Manna\inst{2}\thanks{E-mail: \email{goutammanna.pkc@gmail.com}}}
\shortauthor{Debashis Gangopadhyay and Goutam Manna}

\institute{
\inst{1}Department of Physics, Ramakrishna Mission Vivekananda University, P.O.-Belur Math, Howrah-711202, West Bengal, India.\\
\inst{2}Department of Physics, Prabhat Kumar College, Contai, Purba Medinipur-721401, West Bengal, India.
}


\abstract{
An emergent gravity metric incorporating $k-$essence scalar fields  
$\phi$ having a Born-Infeld type lagrangian is mapped into a metric
whose structure is similar to that  of a 
blackhole of large mass $M$ that has swallowed a global monopole.
However, here the field is not that of a monopole but rather 
that of a $k-$essence scalar field. 
If $\phi_{emergent}$ be solutions of the emergent gravity equations of motion 
under cosmological boundary conditions at $\infty$, 
then for $r\rightarrow\infty$ 
the rescaled field 
$\frac {\phi_{emergent}}{2GM-1}$ has exact correspondence with $\phi$ 
with $\phi(r,t)=\phi_{1}(r)+\phi_{2}(t)$. 
The Hawking temperature of this metric is 
$T_{\mathrm emergent}= \frac{\hbar c^{3}}{8\pi GM k_{\mathrm B}}(1-K)^{2}\equiv  \frac{\hbar}{8\pi GM k_{\mathrm B}}(1-K)^{2}$,
taking the speed of light  $c=1$. Here $K=\dot\phi_{2}^{2}$ is the kinetic energy of the $k-$essence field $\phi$ and $K$ 
is always less than unity, $k_{\mathrm B}$ is the Boltzmann constant. This is phenomenologically interesting 
in the context of Belgiorno {\it et al's} gravitational analogue experiment.}

\def \be{\begin{equation}}
\def \ee{\end{equation}}
\def \ben{\begin{eqnarray}}
\def \een{\end{eqnarray}}

\begin{document}

\maketitle

\section{Introduction}
Actions with non-canonical kinetic terms are  
candidates for dark matter and dark energy.
An  approach to understand their origins 
involves setting up lagrangians for  
$k-$essence scalar fields $\phi$
in a Friedman-Lemaitre-Robertson-Walker metric with zero curvature constant 
\cite{scherrer}. The lagrangian for $k-$essence scalar fields contain non-canonical 
kinetic terms.The general form for such a lagrangian is assumed to
be a function $F(X)$ with $X={1\over 2}g^{\mu\nu}\nabla_{\mu}\phi\nabla_{\nu}\phi$.
Relevant literature involving 
theories with a non-canonical kinetic term and their subsequent use in 
cosmology, inflation, dark matter, dark energy and strings can be found in 
\cite{born,heisenberg,dirac,armen,gib}. 

The motivation of this work is to show that the Hawking temperature \cite{haw} is modified  
in the presence of dark energy. This is realised in the following scenario:
Dynamical solutions of the $k$-essence equation of 
motion changes the metric for the perturbations 
around these solutions \cite{wald,babi}. The perturbations propagate in an emergent spacetime
with metric $\tilde G^{\mu\nu}$ different from (and also not conformally equivalent to) 
the gravitational metric $g^{\mu\nu}$. $\tilde G^{\mu\nu}$ now depends on $\phi$.
Now, if a global monopole falls into a Schwarzschild 
blackhole the resulting metric is different from the Schwarzschild case and the blackhole 
carries the global monopole charge.
Barriola and Vilenkin obtained solutions for Einstein equations outside the monopole core \cite{bv}.  
We determine the $k-$essence scalar field configurations $\phi$ for which the metric $\tilde G^{\mu\nu}$ 
becomes conformally equivalent to  the Barriola-Vilenkin  metric. 
However, in our case the global monopole charge 
is now replaced by the kinetic energy of the $k-$ essence scalar field $\phi$.
Thus we show that BV-type metrics can also result from $k-$essence theories and 
not necessarily from global monopoles only. It should be mentioned that 
monopoles are akin to topological defects and there exist substantial and well-known 
literature on the subject. Recent interesting developments in this area can be found in   
\cite{baz}.     

So one can calculate the Hawking temperature $T_{\mathrm emergent}$  for such a metric and 
this is obviously different from that of the Schwarzschild case. 
Moreover, if $\phi_{emergent}$ be the solutions of the emergent gravity equations 
of motion for $r\rightarrow\infty$ then the rescaled field 
$\frac {\phi_{emergent}}{2GM-1}$ has exact correspondence with the k-essence scalar 
field $\phi$ configurations for which the BV metric is realised.
$M$ is the mass of the BV blackhole and  $M$ is very large   
i.e.,$M >> \frac {\delta}{G}$ where $\delta \sim \lambda^{-\frac {1}{2}}\eta^{-1}$ is the 
monopole core size. Here the global monopole lagrangian is
$L= \frac {1}{2}\partial_{\mu}\phi^{a}\partial^{\mu}\phi^{a} -\frac {1}{4}\lambda(\phi^{a}\phi^{a}-\eta^{2})^{2}$ 
where $\phi^{a}$ is a triplet of scalar fields ($a=1,2,3$) and the global $O(3)$ 
symmetry is spontaneously broken to $U(1)$. $\eta\sim 10^{16} GeV$ is the typical 
grand unification scale.
This result is phenomenologically interesting in the context of   
Belgiorno {\it et al's} \cite{bel1} demonstration of spontaneous emission of photons in a 
gravitational analogue experiment for Hawking radiation.
Accordingly, the plan of the paper is as folows. In the next section we give a brief review 
of emergent gravity. In the following section we show how an emergent gravity metric 
incorporating $k-$essence scalar 
fields $\phi$ and having a Born-Infeld type lagrangian can be mapped into a Barriola-Vilenkin 
metric and how these field configurations are related to solutions of the emergent 
gravity equations of motion for $r\rightarrow\infty$. 
The Hawking temperature is calculated in the subsequent section.
The phenomenological consequences are then discussed in the context of analogue gravity 
experiments of Belgiorno and the last section is the conclusion.

\section{Emergent Gravity}  
The $k-$essence scalar field $\phi$ minimallly coupled to the gravitational 
field $g_{\mu\nu}$ has action  
\ben
S_{k}[\phi,g_{\mu\nu}]= \int d^{4}x {\sqrt -g} L(X,\phi)
\label{eq:1}
\een
where $X={1\over 2}g^{\mu\nu}\nabla_{\mu}\phi\nabla_{\nu}\phi$.  
The energy momentum tensor is
\ben
T_{\mu\nu}\equiv {2\over \sqrt {-g}}{\delta S_{k}\over \delta g^{\mu\nu}}= L_{X}\nabla_{\mu}\phi\nabla_{\nu}\phi - g_{\mu\nu}L
\label{eq:2}
\een
$L_{\mathrm X}= {dL\over dX},~~ L_{\mathrm XX}= {d^{2}L\over dX^{2}},
~~L_{\mathrm\phi}={dL\over d\phi}$ and  
$\nabla_{\mu}$ is the covariant derivative defined with respect to the metric $g_{\mu\nu}$.
The equation of motion is
\ben
-{1\over \sqrt {-g}}{\delta S_{k}\over \delta \phi}= \tilde G^{\mu\nu}\nabla_{\mu}\nabla_{\nu}\phi +2XL_{X\phi}-L_{\phi}=0
\label{eq:3}
\een
where the effective metric $\tilde G^{\mu\nu}$ is 
\ben
\tilde G^{\mu\nu}\equiv L_{X} g^{\mu\nu} + L_{XX} \nabla ^{\mu}\phi\nabla^{\nu}\phi
\label{eq:4}
\een
and is physically meaningful only when $1+ {2X  L_{XX}\over L_{X}} > 0$.

We first carry out the conformal transformation
$G^{\mu\nu}\equiv {c_{s}\over L_{x}^{2}}\tilde G^{\mu\nu}$, with
$c_s^{2}(X,\phi)\equiv{(1+2X{L_{XX}\over L_{X}})^{-1}}\equiv sound ~ speed $.
Then  the inverse metric of $G^{\mu\nu}$ is  
\ben G_{\mu\nu}={L_{X}\over c_{s}}[g_{\mu\nu}-{c_{s}^{2}}{L_{XX}\over L_{X}}\nabla_{\mu}\phi\nabla_{\nu}\phi] 
\label{eq:5}
\een
A further conformal transformation $\bar G_{\mu\nu}\equiv {c_{s}\over L_{X}}G_{\mu\nu}$ gives
\ben \bar G_{\mu\nu}
={g_{\mu\nu}-{{L_{XX}}\over {L_{X}+2XL_{XX}}}\nabla_{\mu}\phi\nabla_{\nu}\phi}
\label{eq:6}
\een
Note that one must always have $L_{X}\neq 0$ for the sound speed 
$c_{s}^{2}$ to be positive definite and only 
then equations $(1)-(4)$ will be physically meaningful. 
This can be seen as follows.   $L_{X}=0$ implies that $L$ does not 
depend on $X$ so that in equation (\ref{eq:1}),  $L(X,\phi)\equiv L(\phi)$.
So the $k-$essence lagrangian $L$ becomes pure potential and 
the very definition of $k-$essence fields becomes meaningless because 
such fields correspond to lagrangians where the kinetic energy dominates 
over the potential energy. If there are no derivatives of the field in 
the lagrangian then there is no question of identifying a kinetic energy 
part.Also, the very concept of minimally coupling the $k-$essence field $\phi$ to 
the gravitational field $g_{\mu\nu}$ becomes redundant and 
equation (\ref{eq:1}) meaningless and equations (\ref{eq:4}-\ref{eq:6}) ambiguous.

For the non-trivial configurations 
of the $k-$ essence field $\partial_{\mu}\phi\neq 0$ and  $\bar G_{\mu\nu}$ is not conformally 
equivalent to $g_{\mu\nu}$. So this $k-$essence
field has properties different from canonical 
scalar fields defined with $g_{\mu\nu}$ and the local causal 
structure is also different from those defined with $g_{\mu\nu}$.
Further, if $L$ is not an explicit function of $\phi$
then the equation of motion $(3)$ is replaced by ;
\ben
-{1\over \sqrt {-g}}{\delta S_{k}\over \delta \phi}
= \bar G^{\mu\nu}\nabla_{\mu}\nabla_{\nu}\phi=0
\label{eq:7}
\een

We shall take the  Lagrangian as $L=L(X)=1-V\sqrt{1-2X}$. 
This is a particular case of the BI lagrangian 
$L(X,\phi)= 1-V(\phi)\sqrt{1-2X}$ for $V(\phi)=V=constant$
and  $V<<kinetic ~ energy ~ of~\phi$ i.e.$V<< (\dot\phi)^{2}$. This 
is typical for the $k-$essence field where the kinetic energy 
dominates over the potential energy.
Then $c_{s}^{2}(X,\phi)=1-2X$.
For scalar fields $\nabla_{\mu}\phi=\partial_{\mu}\phi$. Thus (\ref{eq:6}) becomes
\ben
\bar G_{\mu\nu}= g_{\mu\nu} - \partial _{\mu}\phi\partial_{\nu}\phi
\label{eq:8}
\een
The rationale of using two conformal transformations now becomes clear.
The first transformation is used to identify the inverse metric $G_{\mu\nu}$.
The second conformal transformation realises the mapping onto the   
metric given in $(8)$ for the lagrangian $L(X)=1 -V\sqrt{1-2X}$.

Consider the second conformal transformation $\bar G_{\mu\nu}\equiv {c_{s}\over L_{X}}G_{\mu\nu}$.
Following \cite{wald} the new Christoffel symbols are related to the old ones by  
$$
\bar\Gamma ^{\alpha}_{\mu\nu} 
=\Gamma ^{\alpha}_{\mu\nu} + (1-2X)^{-1/2}G^{\alpha\gamma}[G_{\mu\gamma}\partial_{\nu}(1-2X)^{1/2}$$
$$+G_{\nu\gamma}\partial_{\mu}(1-2X)^{1/2}-G_{\mu\nu}\partial_{\gamma}(1-2X)^{1/2}]$$
$$=\Gamma ^{\alpha}_{\mu\nu} -\frac {1}{2(1-2X)}[\delta^{\alpha}_{\mu}\partial_{\nu}X
+ \delta^{\alpha}_{\nu}\partial_{\mu}X]$$
Note that the second term on the right hand side is symmetric under exchange of $\mu$ and $\nu$ 
so that the symmetry of $\bar\Gamma$ is maintained.The second term has its origin solely to the 
$k-$essence lagrangian and this additional term signifies additional interactions (forces). 
The geodesic equation in terms of the new Christoffel connections  $\bar\Gamma$ now becomes 
$$\frac {d^{2}x^{\alpha}}{d\tau^{2}} +  \bar\Gamma ^{\alpha}_{\mu\nu}\frac {dx^{\mu}}{d\tau}\frac {dx^{\nu}}{d\tau}=0$$
\section{Mapping on to the  Barriola-Vilenkin type metric }
Taking the  gravitational metric $g_{\mu\nu}$ to be Schwarzschild,
$\partial_{0}\phi\equiv\dot\phi$, $\partial_{r}\phi\equiv\phi '$ and assuming 
that the $k-$ essence field $\phi (r,t)$ is spherically symmetric one has 
\ben
\bar G_{00}= g_{00} - (\partial _{0}\phi)^{2}=1-2GM/r - \dot\phi ^{2}\nonumber\\
\bar G_{11}= g_{11} - (\partial _{r}\phi)^{2}= -(1-2GM/r)^{-1} - (\phi ') ^{2} \nonumber\\
\bar G_{22}= g_{22}=-r^{2}\nonumber\\ 
\bar G_{33}= g_{33}=-r^{2}sin^{2}\theta\nonumber\\
\bar G_{01}=\bar G_{10}=-\dot\phi\phi ' 
\label{eq:9}
\een
For the Schwarzschild metric,
$g_{00}=(1-2GM/r);g_{11}=-(1-2GM/r)^{-1};
g_{22}=-r^{2}; g_{33}=-r^{2}sin^{2}\theta; g_{ij} (i\neq j)=0$.
So the emergent gravity line element becomes
\ben
ds^{2}=(1-2GM/r - \dot\phi ^{2})dt^{2}\nonumber\\-((1-2GM/r)^{-1} + (\phi ') ^{2})dr^{2}\nonumber\\
-2\dot\phi\phi 'dtdr-r^{2}d\Omega^{2}\label{eq:10}
\een
where $d\Omega^{2}=d\theta^{2}+sin^{2}\theta d\Phi^{2}$.

Making a co-ordinate transformation from $(t,r,\theta,\phi)$ to 
$(\omega,r,\theta,\phi)$ such that (\cite{wein}):
\ben
d\omega=dt-({{\dot\phi\phi '}\over{1-2GM/r - \dot\phi ^{2}}})dr\label{eq:11}
\een
Then (\ref{eq:10}) becomes
\ben
ds^{2}=(1-2GM/r - \dot\phi ^{2})d\omega^{2}\nonumber\\
-[{{(\dot\phi\phi ')^{2}}\over{(1-2GM/r-\dot\phi^{2})}}+{1\over(1-2GM/r)}+{(\phi ') ^{2}}]dr^{2}\nonumber\\
-r^{2}d\Omega^{2}\nonumber\\
\label{eq:12}
\een
(\ref{eq:12}) will be a blackhole metric if $\bar G_{00}= \bar G_{11}^{-1}$ , i.e.
\ben
\dot\phi^{2}=(\phi')^{2}(1-2GM/r)^{2}
\label{eq:13}
\een
Let us assume a solution to (\ref{eq:13}) of the form 
$\phi(r,t)=\phi_{1}(r)+\phi_{2}(t)$.
Then (\ref{eq:13}) reduces to 
\ben
\dot\phi_{2}^{2}=(\phi_{1}')^{2}(1-2GM/r)^{2}= K
\label{eq:14}
\een 
$K(\neq 0)$ is a constant ($K\neq 0$ means $k-$essence field will have {\it non-zero} kinetic energy).
The solution to (\ref{eq:14}) 
\ben
\phi(r,t)=\phi_{1}(r)+\phi_{2}(t)\nonumber\\
=\sqrt{K}[r+2GM~ln(r-2GM)]+\sqrt{K}t
\label{eq:15}
\een
with  $\phi_{1}(r)=\sqrt{K}[r+2GM ln(r-2GM)] ; \phi_{2}(t)=\sqrt{K}t$,
and we have taken an arbitrary integration constant to be zero.
So the line element (\ref{eq:12}) reduces to 
\ben
ds^{2}=(1-{2GM\over r}-K)d\omega^{2}\nonumber\\
-{1\over(1-{2GM\over r}-K)}dr^{2}-r^{2}d\Omega^{2}\nonumber\\
\label{eq:16}
\een
and this is the Barriola-Vilenkin blackhole which represents the situation where a global monopole 
carrying charge $K=\dot\phi_{2}^{2}=constant$ has fallen into a Schwarzschild blackhole. 
{\it It should be noted that $K$ has to be always less than unity because if $K$ is greater 
than unity the signature of the metric $(16)$ becomes ill defined.} This is easily seen : for $K>1$, 
$\bar G_{00}$ is negative while $\bar G_{11}$ is positive. However, it should also be 
noted that $K >> V$. This is a requirement for $k-$essence fields where the 
kinetic energy dominates over the potential energy. Therefore, we have  
$K < 1$ and $V << K$.

So the metric components are
\ben
\bar G_{00}= g_{00} - (\partial _{0}\phi)^{2}=(1-2GM/r - K)\nonumber\\
\bar G_{11}= g_{11} - (\partial _{r}\phi)^{2}=-(1-2GM/r-K)^{-1}\nonumber\\
\bar G_{22}= g_{22}=-r^{2}~~;~~\bar G_{33}= g_{33}=-r^{2}sin^{2}\theta\nonumber\\
\label{eq:17}
\een
In the context of global monopoles \cite{bv}, the 
above metric has been shown to satisfy 
the Einstein field equations. Thus  we have shown that this metric can also 
arise from an emergent gravity scenario with $k-$essence scalar fields and the 
global monopole charge is now replaced by the constant kinetic energy of the 
$k-$essence field.

Now it is also known that the solutions to the emergent gravity equations of motion
(\ref{eq:7}) for the scalar field under cosmological boundary conditions are given by \cite{babi}
$\phi_{emergent}(t,r)=const. [t+r+2GM~ln~|\frac{r}{2GM} -1|+2GM\int^{r} F(r')dr']$ where the function 
$F(r)=\frac {r}{r-2GM}[\sqrt{\frac{Ar-2GM}{A^{4}r^{4}(r-2GM)+(A-1)r}}-1]$ where $A$ is a constant.  
Substituting this solution in (\ref{eq:13}) and taking the limit $r\rightarrow\infty$
and ignoring terms of $O(\frac{1}{r^{2}})$ and higher gives 
\ben
lim_{r\rightarrow\infty}(\phi_{emergent}')^{2}(1-2GM/r)^{2} = K(2GM -1)^{2}
\label{eq:18}
\een
Therefore for $r\rightarrow\infty$ ,{\it the rescaled field} $\frac {\phi_{emergent}(t,r)}{(2GM-1)}$ 
{\it has exact correspondence with the $k-$essence scalar field $\phi$ which satisfies the blackhole 
metric condition} (\ref{eq:13}).
\section {Hawking Temperature} 
Let us calculate the Hawking temperature for this 
metric (\ref{eq:16}) using the tunnelling formalism as outlined in \cite{mitra} which  
corrects for the factor of two in the Hawking temperature as often mentioned, 
(e.g. in \cite{akh}).
Going over to the Eddington-Finkelstein coordinates $(v,r,\theta,\phi)$
\ben
v=\omega+r^{*}~~;~~u=\omega-r^{*}\nonumber\\
r^{*}= {r\over 1-K} + {2GM\over (1-K)^{2}}~ln~[(1-K)r-2GM]
\label{eq:19}
\een
we get
\ben
ds^{2}= (1-K-2GM/r)dv ^{2} - 2 dv dr - r^{2}d\Omega ^{2}
\label{eq:20}
\een
By analogy with the  Schwarzschild case a massless particle 
in the  background  of $\bar G_{\mu\nu}$ is described
by the Klein-Gordon equation
\ben
\hbar^2(-\bar G)^{-1/2}\partial_\mu( \bar G ^{\mu\nu}(-\bar G)^{1/2}\partial_\nu\Psi)=0.
\label{eq:21}
\een
One expands
\ben
\Psi=\exp(-{i\over\hbar}S+...)
\label{eq:22}
\een

\ben
\bar G^{\mu\nu}\partial_\mu S\partial_\nu S=0.
\label{eq:23}
\een
Using separation of variables in the form
\ben
S=Et+S_0(r),
\label{eq:24}
\een
and proceeding exactly as outlined in \cite{mitra}, the Hawking temperature
for this metric is obtained as ($\dot\phi_{2}^{2}=constant=K$)
\ben
T_{\mathrm emergent}= {\hbar c^{3}(1-K)^{2}\over 8\pi GMk_{B}}\nonumber\\
= T_{\mathrm H} (1-K)^{2}
\label{eq:25}
\een
where $T_{\mathrm H}={\hbar c^{3}\over 8\pi GMk_{B}}$ is the usual Hawking temperature.
So $T_{\mathrm emergent}$ is less than the usual Hawking temperature 
for Schwarzschild black hole as $K < 1$.

Let us recollect what we have done so far. We have considered a $k-$essence scalar field $\phi$ 
(with a non-canonical Born-Infeld type lagrangian with potential $V(\phi)=const.=V$ i.e.
$L=1-V(\phi)\sqrt {1-2X}\equiv 1-V\sqrt {1-2X}$) minimally coupled to the gravitational metric 
$g_{\mu\nu}$ in Schwarzschild spacetime. We then obtain  
the equation of motion for $\phi$, equation (\ref{eq:7}), in the  ''effective'' metric 
$\bar G_{\mu\nu}= g_{\mu\nu}-\partial_{\mu}\phi\partial_{\nu}\phi$.

We then impose the conditions for a blackhole metric to obtain the configurations of the 
$k-$essence field $\phi$ that will give a blackhole. It turns out that one possible scenario 
is a Barriola-Vilenkin type blackhole where the global monopole charge is now replaced by a constant 
kinetic energy of the scalar field $\phi$. This kinetic energy $K < 1$ in order to preserve the 
consistency of the signature of the metric. 
By construction the potential energy $V$ is a constant and $V<<K$ because this is a basic requirement 
for $k-$essence fields. Therefore, the
total energy is always a constant and although the obtained field configurations are linear 
in time there should not be any instability.Moreover, the lagrangian does not depend on the fields $\phi$ 
explicitly. The dependence is only through derivatives of the field.
\section{Phenomenological consequences in analogue gravity experiments}
We now discuss how our results can be made to correspond to scenarios 
in analogue gravity experiments similar to  Belgiorno {\it et al} \cite{bel1}. Such an 
experiment may help in  distinguishing between a Schwarzschild blackhole analogue  
and the blackhole analogue described by the metric (\ref{eq:16}) and (\ref{eq:17}).

First let us briefly discuss Belgiorno {\it et al's} experiment. 
They used ultrashort laser pulse filaments to create a travelling refractive index 
perturbation (RIP) in fused silica glass and reported experimental evidence of photon emission 
that bears the characteristics of Hawking radiation  and is distinguishable and thus separate 
from other known photon emission mechanisms. They interpreted this emission as an indication 
of Hawking radiation induced by the analogue event horizon. 

They also have a complete description of the event horizon associated to the RIP and can 
calculate a blackbody temperature of the emitted photons in the laboratory reference frame
\cite{phil,bel2}. However, Belgiorno {\it et al} also pointed out that the dielectric medium 
in which the RIP is created will always be dominated by optical dispersion and therefore the 
spectrum will not be that of a perfect blackbody and that in any case only a limited spectral 
portion of the full  spectrum will be observable. To show this last point they described the 
RIP  as a perturbation induced by the laser pulse on top of a uniform, dispersive background 
refractive index $n_0$, i.e. $n(z,t,\omega)=n_0(\omega)+\delta n f(z-vt)$, where $\omega$ is 
the optical frequency, $f(z-vt)$ is a function bounded by 0 and 1, that describes the shape 
of the laser pulse. In the reference frame co-moving at velocity $v$ with the RIP, the 
event horizon in a 2D geometry is defined by  $c/v=n$ which admits solutions only for RIP 
velocities satisfying the inequality \cite{bel2}:
$\frac{1}{n_0(\omega)+\delta n}<\frac{v}{c}<\frac{1}{n_0(\omega)}$.
This predicts an emission spectrum with well-defined boundaries and it is precisely this feature 
of the spectral emission that is peculiar to analogue Hawking radiation.
In the experiment a clear photon emission was registered in the  wavelength window  predicted by 
the last inequality, the emitted radiation was unpolarized, and
the emission bandwidth increased with the 
input energy. The Bessel pulse
intensity evolution along the propagation direction $z$ was estimated analytically from the 
input energy. By fitting the measured spectra with Gaussian functions  the bandwidth 
was estimated as a function of input energy and Bessel intensity. Using the fused 
silica dispersion relation the authors obtained the bandwidth and the $\delta n$ as a function 
of input energy and Bessel pulse peak intensity (at $z=1$ cm where measurements were performed). 
There was a clear linear 
dependence which was in qualitative agreement with the fact that the emission bandwidth 
was predicted to depend on $\delta n$ which in turn is a linear function 
$\delta n = n_2I$ of the pulse intensity $I$.
The slope of the linear fit was in good agreement with the tabulated 
value \cite{desalvo,couairon}. Therefore there is also an agreement at the quantitative 
level between the measurements and the model based on Hawking-like radiation emission.

{\it In this context, we propose that a different Hawking temperature should give a different set 
of values for the above mentioned phenomenological parameters i.e. the slope $n_{2}$ and also 
the inequality $\frac{1}{n_0(\omega)+\delta n}<\frac{v}{c}<\frac{1}{n_0(\omega)}$ etc. because 
the intensity of photon emission should depend on the relevant blackbody temperature.} 
Therefore, Belgiorno {\it et al's} gravitational analogue experiment has scope of being 
further enhanced into testing the existence of  other cosmological entities like
dark energy. In order to include the effects of dark energy, the RIP method must be 
accordingly modified. At a basic level this means that the effect of the presence 
of the constant $K$ in the metric must be included. For example, if we consider 
the Schwarzschild metric, the Belgiorno {\it et al} 
experiment currently has $K=0$. So the experimental situation has to move over 
to a scenario which can mimic $K\neq 0$ 
i.e. $0 < K < 1$. 

Here some aspects need to be clarified.First, note that in $(16)$ 
if we take $M\sim M_{\mathrm monopole core}$ i.e. $M$ is very small 
and negligible we have
$ds^{2}=(1-K)d\omega^{2}-{1\over(1-K)}dr^{2}-r^{2}d\Omega^{2}$ and 
rescaling $r$ and $\omega$ one has 
$ds^{2}=d\omega^{2}-dr^{2}-(1-K) r^{2}d\Omega^{2}$ and this is the 
metric of the  global BV monopole and describes a space 
with a deficit solid angle i.e. the area of a sphere of radius 
$r$ is not $4\pi r^{2}$ but $(1-K) 4\pi r^{2}$ , $K <1$. Such spaces 
are not asymptotically flat, but asymptotically bound.

Secondly, we are dealing with the BV blackhole i.e.
$M >> M_{\mathrm monopole core}$ i.e. $2GM / r = \frac {2M}{ {r\over G}} $ is not 
negligible i.e. for  $r\sim\delta$ where $\delta$ is the monopole core size. Here 
also for $r\rightarrow\infty$ the metric $(16)$ is not strictly asymptotically flat 
owing to the presence of $K$.    
 
If the Schwarzschild metric is the reference then
one has to move over from an asymptotically 
flat metric to one which is not exactly asymptotically flat but rather 
asymptotically bound. This aspect can be incorporated into the  
Belgiorno {\it et al} analogue gravity experiment in the following way in order for the 
results to be compatible to the existence of an analogue event horizon corresponding to a 
Hawking temperatute $T_{\mathrm emergent}$.
Here we draw heavily from Reference \cite{bel2}. Consider the wave equation  
for a perturbation of a full nonlinear electric field propagating 
in a nonlinear Kerr medium where for simplicity the electric field has  been replaced by 
a scalar field $\phi$, (equation $(1)$ of Reference \cite{bel2}) :
$$\frac {n^{2}(x_{l}-vt_{l})}{c^{2}} \partial _{t_{l}} ^{2} \Phi  
-\partial _{x_{l}} ^{2} \Phi - \partial_{y}^{2} \Phi - \partial _{z} ^{2} \Phi = 0 $$ 
where all coordinates are in the lab frame. The suffix $l$ is omitted from $y,z$ 
because they are not involved in the boost relating the lab frame with the pulse frame.
$n(x_{l}-vt_{l})$ is the refractive index that accounts for the propagating 
RIP in the dielectric. The RIP is propagating with a constant velocity $v$.
The analogue Hawking temperature (equation $(13)$ 
of Reference \cite{bel2}) for the blackhole horizon ($x_{+}$) is 
$$T_{+}=\frac {\gamma ^{2}v^{2}\hbar} {2\pi k_{B}c} |\frac {dn}{dx}|_{x_{+}}$$
where $\gamma$ is the boost, $v$ is the constant velocity of the RIP, $k_{B}$ is the 
Boltzmann constant and $x_{+}$ denotes the blackhole horizon. Note that in order 
to have $T_{emergent}=(1-K)^{2}T_{H}$ one can have either of the following scenarios:

{\it Case 1:}

Realise an experimental situation where 
$$\gamma ^{2}\rightarrow \gamma _{1} ^{2}= (1-K)^{2}\gamma ^{2}$$
This would imply
$$v _{1} ^{2}  = \frac {v^{2} - c^{2}K(2-K)}{(1-K)^{2}}$$
As $v _{1} ^{2}$ must be positive, one should ensure that 
$$K(2-K) < \frac {v^{2}}{c^{2}}$$
All this should also be made consistent with 
$$\frac {c}{v_{1}}= n_{1} |_{x_{+}}= n_{10} + k _{1}\eta _{1}$$
where the parameter $\eta _{1}<<1$ and $k_{1}$ is the normalised intensity of the pulse at the blackhole horizon taking values in 
the interval $(0,1)$.

{\it Case 2:}

Realise an experimental situation with 
$$n\rightarrow n_{2}=(1-K)^{2} n$$
Obviously the RIP propagation velocity $v$ must be changed to some 
new value $v_{2}$ together with a new consistency condition 
$$\frac {c}{v_{2}}= n_{2} |_{x_{+}}= n_{20} + k _{2}\eta _{2} $$ 

{\it Therefore , a realisation of photon emission (similar to Belgiorno
{\it et al's} original experiments) corresponding to  a blackbody 
temperature $T_{\mathrm emergent}$ in any of the  
above described two scenarios will be  
compatible with a theory of $k-$essence fields in an emergent 
gravity metric in the same way as the original Belgiorno 
experiment, though not proving the existence of blackhole horizons , 
however , proves the 
relation between blackhole horizons and the Hawking radiation.}
\section{Conclusion}
We have described here a situation where 
a BV-type metric can also result from scenarios with 
$k-$essence scalar fields. The Hawking temperature for such a metric has  
a dark energy component and we have indicated a phenomenological scenario
(in the context of Belgiorno {\it el's} gravitational analogue experiment 
\cite{bel1,bel2}) where there is  scope for this to be tested. 

Our work is in the domain of 
an emergent gravity metric $\bar G^{\mu\nu}$ incorporating $k-$essence scalar fields  
$\phi$ having a Born-Infeld type lagrangian.
We first determine the scalar field configurations for which $\bar G^{\mu\nu}$  
is mapped into a BV type metric. We then show that if \revision {$\phi_{emergent}$} be solutions 
of the emergent gravity equations of motion 
under cosmological boundary conditions at $\infty$, 
then for $r\rightarrow\infty$ 
the rescaled field 
$\frac {\phi_{emergent}}{2GM-1}$ has exact correspondence with $\phi$ 
with $\phi(r,t)=\phi_{1}(r)+\phi_{2}(t)$.
The Hawking temperature of the resulting BV-type metric is found to be 
$T_{\mathrm emergent}= (1-K)^{2}T_{H}$.
Here $K=\dot\phi_{2}^{2}$ is the kinetic energy of the $k-$essence field $\phi$ and $K$ 
is always less than unity. We have then indicated why  certain phenemenological 
parameters in Belgiorno's 
analogue gravity experiment will be modified because of the difference  
of the Schwarzschild metric from that of a BV-type metric. Finally, we 
described certain analogue gravity experimental situations with which  
the compatibility of our theoretical considerations may be tested.

The authors would like to thank the referees for illuminating suggestions to 
improve the manuscript.

\end{document}